\documentclass[twocolumn,showpacs,titlepage,aps,prd,letter,
tightenlines,amsmath,byrevtex,nofootinbib,floatfix]{revtex4}

\begin{document}
\title{Explanation for the low flux of high energy
            astrophysical muon neutrinos}

\author{Sandip Pakvasa}

\affiliation{Department of Physics and Astronomy,
                 University of Hawaii,
                     Honolulu, HI 96822
                        U.S.A.}

\author{Anjan Joshipura and Subhendra Mohanty}

\affiliation{Physical Research Laboratory, Ahmedabad, India 380009}

\date{\today}

\begin{abstract}

There has been some concern about the unexpected paucity of cosmic high
energy muon neutrinos in detectors probing the energy region beyond 1 PeV.
As a possible solution we consider the possibility that some exotic neutrino
property is  responsible for reducing the muon neutrino flux
at high energies from distant sources; specifically, we consider:
(i) neutrino decay and
(ii) neutrinos being pseudo-Dirac particles. This would provide a
mechanism for
the reduction of high energy muon events in the IceCube detector, for example.
\end{abstract}

\maketitle


            The most recent data from the IceCube
collaboration\cite{abbasi} place stringent limits on the muon neutrino
flux  at high energies from astrophysical sources.
The new limits appear to put severe bounds on models of neutrino
production in GRB's and AGN's\cite{waxman}. Similarly, other experiments
probing the ultra-high-energy regime, such as ANITA\cite{vieregg} and AUGER\cite{auger}
have not seen any evidence of long anticipated cosmic neutrinos. It should be noted that very recently there have been re-evaluations of the expected neutrino fluxes from GRB's, especially following the stringent upper limits from IceCube \cite{abbasi}.
It has been pointed out \cite{LiZh,Hummer:2011ms} that IceCube  \cite{abbasi} calculation of the WB neutrino flux from the observed gamma ray flux may have been an overestimation by as much as a factor of 5.
So the discrepancy may not be that dire, yet; but the possibility remains that as the bounds get tighter with future observations, the Waxman-Bahcall models \cite{waxman} will be challenged.
In such an eventuality, we would like to offer in this letter the possibility
of other causes for the smallness of the muon neutrino flux, which arise from  neutrino properties. We note that there are alternative astrophysical models ( see \cite{Gao:2012ay,Baerwald:2013pu} and references therein)  which predict a lower neutrino flux compared to the Waxman-Bahcall models \cite{waxman}.


In this note we would like to raise the
 possibility that these severe bounds are illusory because the small
flux may be due to depletion of muon neutrinos which in turn is
 caused by neutrino properties
We consider two possible  scenarios. One is that
neutrino decay is responsible for depletion
of muon-neutrinos and the other is that neutrinos are pseudo-Dirac particles and there is leakage into the sterile components of the pseudo-Dirac particles. Both of these were
considered  almost ten years ago\cite{beacom,beacom1}, but the focus then was on
the modification of the flavor mix from the canonical 1:1:1
as expected from conventional flavor oscillations with the
known neutrino mixings\cite{learned}.

In the following, we describe both possibilities. To be definite, we are considering neutrino
energies in the vicinity of order of a PeV, and the
distances from the sources of order of hundreds of
mega-parsecs. In principle, when the distances become large enough, the cosmological red
shift becomes important, and the travel distance L is
limited; these effects were discussed some time ago in
ref.\cite{beacom1,weiler1} and more recently in ref.\cite{baerwald}
and ref.\cite{esmaili}.

Of course, because of the uncertainty in
predicting fluxes, we do not know precisely what amount of depletion is
needed.  But the scenarios we suggest below can provide a wide range of
suppression ranging from none to an order of magnitude.
\begin{flushleft}
Neutrino Decay:
\end{flushleft}

We consider here scenarios with three light neutrinos and assume that the source distances are large
enough so that two of the three mass eigenstates,
specifically  $\nu_2$ and $\nu_3$ have decayed away completely. If the neutrino masses are quasi-degenerate, that is the masses of $\nu_2$  and $\nu_3$ are close to that of $\nu_1$, then the daughter neutrino, $\nu_1$ carries most of the energy of the parent, and so contributes to the flux at that energy; in this case even though the final state is pure $\nu_1$, there is not much depletion. So for our purpose here, the preferred mass spectrum is quasi-hierarchical, namely $m_2$ and $m_3$ much larger than $m_1$; in this case the daughter neutrino energy is much lower than the parent and the final $\nu_1$ does not contribute to the flux at that energy and can be counted out. This is discussed in detail in several papers, especially clearly in ref.\cite{Beacom-Bell}. This
means that the exponential decay factor exp $(-L/\gamma c \tau)$ is
negligibly small for them.  Since distances to GRB's are of
order of 100's of MPc, for energies in the PeV range,
$\frac{L}{\gamma c \tau} = \frac{L}{E} \left (
\frac{mc^2}{c \tau} \right ) \gg 1$ corresponds to $\tau/m <
10^{3} sec/eV$ where $\tau$ is the rest frame lifetime.
A lower bound on the lifetime follows from the BBN (Big Bang Nucleosynthesis). If the standard  picture is to remain intact then all three neutrinos must be present and in equilibrium in the BBN era so that
the crucial n/p ratio and the nuclear abundances as obtained in standard picture remains unaffected. This puts a lower bound of $\frac{\tau}{m} E>1$ sec on the neutrino lifetime
with $E\sim $ MeV. These considerations restrict the allowed window of lifetime in the range
\begin{equation}
 \label{range}
10^{-6} sec./eV \leq \frac{\tau}{m}\leq 10^3  sec./eV~.
\end{equation}

As for the neutrino decay modes, we know the following.  The radiative
decays such as $\nu_i \rightarrow v_j + \gamma$ are severely constrained
by their contribution to $\nu + e \rightarrow e + \nu'$ and from the
current bounds on such contributions the radiative decay lifetime must
satisfy \cite{pakvasa}.
\begin{equation}
\tau_i/m_i \ \ > 10^{17} \ sec./eV.
\end{equation}
The three-body invisible decay mode
\begin{equation}
\nu_i \rightarrow \nu_j \ + \nu \bar{\nu}
\end{equation}
is constrained by BBN and the deviation of the invisible width of Z from
the expected value (with three neutrinos) in SM \cite{bilenky};
and is given by
\begin{equation}
\tau_i/m_i > 10^{28} sec./e V
\end{equation}

The kinds of decay models possible are quite restricted. Models where the coupling is chirality conserving (e.g. into a light vector
boson or into a scalar boson with a derivative coupling),
would by $SU(2)_L \times U(1)$
symmetry  lead to flavor changing decays of charged leptons at the same strength.  The severe bounds on flavor changing decays of $\mu$ and $\tau$ into invisible two body modes lead to limits on lifetimes of $\nu_2$ and $\nu_3$ of order of $\tau > 10^{20}$ sec  \cite{Jodido},
and so such decays are ruled out.
Hence, the only neutrino decay modes which can be relevant for the short
lifetimes of interest here
are helicity changing  decays into a neutrino and a light
boson, as discussed in ref.\cite{beacom,pakvasa}. The current limits
on the lifetimes of the three mass eigenstates
are as follows. The most stringent is on that of $\nu_1$, from the
observation of neutrinos from SN1987A as being about $\tau_1/m_1 >
10^5 s/eV$\cite{SN1987A}.    The limits on the other two mass eigenstates are:
$\tau_2/m_2 > 10^{-4} s/eV$ from the solar neutrino observations\cite{Beacom-Bell,jmm} and
$\tau_3/m_3 > 10^{-10} s/eV$ from the atmospheric neutrino observations\cite{gonzalez}.
Obviously, the limits on the lifetimes of $\nu_2$ and $\nu_3$
are quite weak.

In the picture adopted here, all the neutrinos originating from GRBs reach the earth as pure $\nu_1$ whose flavour
content is  $e:\mu:\tau =
|U_{e1}|^2:|U_{\mu 1}|^2:|U_{\tau 1}|^2$ as observed
long ago\cite{pakvasa1}.
If we insert the current best fit values \cite{schwetz} for the Maki-Nakagawa-Sakata-Pontecorvo (MNSP) \cite{MNSP}  neutrino mixing matrix  elements, we find
that $|U_{\mu 1}|^2$ ranges between 0.1 and
0.3 with a
central value of about 0.16. [The unknown value of the CP violating
  phase $\delta$ in the MNSP mixing matrix  
determines the precise value].
This is a suppression beyond the factor of two due to the standard flavor oscillations. Thus, a suppression of the muon neutrino flux by
an order of magnitude is easily achieved. Since the value of $|U_{e1}|^2$ is
between 0.65 and 0.72, the $\nu_e$ flux
is not affected much by the decays of $\nu_2$ and $\nu_3$ .
We note that the flux ratio of
$\nu_e$ to $\nu_\mu$ is between 2.5 and 8 with a central value of about 4, depending on the
value of the phase $\delta$.
We have discussed the most favorable scenario for $\nu_\mu$ flux reduction by assuming (i) normal hierarchy, because in the inverted hierarchy the decay of $\nu_1$ has strong limits from SN1987a so only $\nu_2$ can decay into $\nu_3$ but in that case we do not achieve any suppression of $\nu_\mu$ and (ii) hierarchial masses, namely $m_2,m_3 \gg m_1$ ; otherwise if the masses are degenerate, the energy of the decaying and daughter neutrino are the same and  even though the flavor ratio $\nu_e/\nu_\mu$ is large there is not much suppression of $\nu_\mu$ flux because of enhancement of the $\nu_1$ flux from the decay.

The invisible decays $\nu_{2,3}\rightarrow \nu_1+J$ arise naturally in Majoron models with $J$ identified  with the massless Majoron  arising from the spontaneous breaking of total lepton number or some combination of $L_i$, $i=e,\mu,\tau $. These models fall in two main categories: triplet majoron models \cite{tripletmajoron}  with a low scale lepton number violation and singlet models \cite{singletmajoron}  with lepton number typically broken at high scale. The former class of models give a  large contribution to the invisible decay width of the $Z$ boson and are ruled out. The singlet majoron models are consistent with the $Z$ decay width but mixing of  Majorn  with the doublet Higgs in this case lead to rapid energy loss from stars  through majoron emission. This can be prevented  if lepton number breaking occurs at a high scale,  (typically $> 10^7$ GeV). It is however possible to consider hybrid models in which Majoron is a combination of the $SU(2)_L$ doublet, triplet and singlet. Such models allow low lepton number breaking
scale and can be made consistent with the existing experimental constraints  \cite{lowscale}.

The Majoron couplings to neutrinos is  flavour diagonal  in simplest triplet model \cite{tripletmajoron} and are nearly so in singlet majoron \cite{singletmajoron}  models. Both these do not allow short neutrino lifetime as required in eq.(1) but such life times  can be achieved by allowing majoron to be associated with a spontaneous breaking of some combination of lepton numbers \cite{decay} and may also need extension of the simplest model.
Denote the coupling relevant to decay as
\begin{equation}
g_{1a}\frac{m_a}{f}\overline{\nu}_1\gamma_5\nu_a J~,
\end{equation}
where $m_a$, $a=2,3$ is the mass of the decaying neutrino, $f$ is the symmetry breaking scale and $g_{1a}$ is a model dependent
overall coupling. The allowed window for the life time as given in eq.(\ref{range}) constrains the symmetry breaking scale to
lie in the range
\begin{equation}
 \label{franges}
8 \,{\rm eV} \leq\frac{f}{g_{1a}}\leq 0.15\, {\rm MeV}
\end{equation}
for normal hierarchy with $m_a\sim 0.05$ eV. This in  particular rules out models with lepton number broken at very high scale but  hybrid models \cite{lowscale,decay}  with a low $f$ are still allowed.

It has been pointed out \cite{Hannestad:2005ex} that neutrino interactions with a light scalar can make the neutrino
fluid tightly coupled at the time of photon decoupling (when $T_\gamma=0.256$ eV). If neutrinos do not free-stream after photons decouple then they can be a source for photon perturbations which would be observable in CMB anisotropy data.
If all neutrinos are assumed to be free-steaming during decoupling then neutrino-Majoron coupling and hence the scale $f$ is $\frac{f}{g_{1a}}> 10^{11} m_a\sim 5 {\rm GeV}$  which would rule out decay of PeV neutrinos over cosmological distances of 100 Mpc.
However it has been shown in \cite{Bell:2005dr} that the CMB data does not preclude one or even two neutrino species from being strongly coupled $( g_{1a}\,m_a/f) > 10^{-7}$ and this still keeps the possibility of UHE neutrino decay viable. The recent Planck data  may be able to
put stronger constraints on the number of tightly coupled neutrinos at decoupling and rule out the possibility of UHE neutrino invisible decays \cite{Friedland:2007vv}.

Amongst other consequences, the neutrino counting in early universe is
modified  from a count of 3 to $3+4/7$ due to the extra bosonic degree
of freedom. This is consistent with most recent cosmological bounds\cite{WMAP}.

The bottom line is that if neutrinos decay, substantial reduction in $\nu_\mu$
fluxes is  possible, and consistent with $\nu_1$ being the lightest mass eigenstate.

\begin{flushleft}
Pseudo-Dirac Neutrinos:
\end{flushleft}
If each of the three neutrino mass eigenstates is actually a doublet
with very small mass difference (smaller than $10^{-6} eV)$, then there
are no current experiments that could have detected this.  Such a
possibility was raised long ago \cite{wolfenstein}. It turns out that the
only way to detect such small mass differences in the range $(10^{-12} eV^2 > \delta
m^2 > 10^{-18} eV^2)$ is by measuring flavor mixes of the high energy
neutrinos from cosmic sources.

Let $(\nu_1^+, \nu_2^+, \nu^+_3; \nu^-_1 \nu^-_2, \nu^-_3$ denote the
six mass eigenstates where $\nu^+$ and $\nu^-$ are a nearly degenerate
pair.  A 6$\times$6 mixing matrix rotates the mass basis into the flavor basis
$(\nu_e, \nu_\mu, \nu_\tau; \nu'_e, \nu'_\mu, \nu'_\tau)$.  In general, for
six Majorana neutrino, there would be fifteen rotation angles and fifteen
phases .  However, for pseudo-Dirac neutrinos, Kobayashi and
Lim \cite{kobayashi} have given an elegant proof that the 6x6 matrix
$V_{KL}$ takes the very simple form
\begin{eqnarray}
V_{KL} \ = \left (
\begin{array}{cc}
U & 0\\
0 & U_R
\end{array}
\right) .
\left (
\begin{array}{cc}
V_1 & iV_1 \\
V_2 & -i V_2
\end{array}
\right ) .
\end{eqnarray}
where the 3$\times$3 matrix $U$ is just the usual mixing matrix; the 3$\times$3 matrix $U_R$ is an
unknown unitary matrix and $V_1$ and $V_2$ are the diagonal matrices
$V_1 = diag (1,1,1) / \sqrt{2},$ and $V_2 = diag (e^{-i \phi 1}, e ^{-i
  \phi 2}, e^{-i \phi 3})$ / $\sqrt{2}$ with the $\phi_i$ being
arbitrary phases.
A very similar mass spectrum can be produced
in the mirror model \cite{mirror-models}.

As a result, the three active neutrino states are described in terms of
the six mass eigenstates as:
\begin{equation}
\nu_{\alpha L} = U_{\alpha j} \frac{1}{\sqrt{2}} (\nu^+_j \ + i \nu^-_j
).
\end{equation}

The nontrivial matrices $U_R$ and $V_2$ are not accessible to active
flavor measurements.  The flavor conversion probability can thus be
expressed as
\begin{eqnarray}
P_{\alpha \beta} = \frac{1}{4} \left |
\sum _{j=1}^{3} U_{\alpha j} \left \{ e^{i (m_j^+)^2 l/2E} +
e^{i(m^-_j)^2 l/2E} \right \} U^*_{\beta j}  \right |^2
\end{eqnarray}

In the description of the three active
neutrinos, the only new parameters are the three pseudo-Dirac mass
differences, $\delta m^2_j = (m_j^+)^2 - (m_j^-)^2$.  In the limit that
they are negligible, the oscillation formulas reduce to the standard
ones and there is no way to discern the pseudo-Dirac nature of the
neutrinos.

Incidentally, the effective mass for neutrino-less double beta decay is
given by
\begin{equation}
\langle m \rangle_{eff} = \frac{1}{2} \sum_j \ U^2_{ej} (m^+_j - m^-_j) =
\frac{1}{2} \sum_{j} \ U^2_{ej} \
\frac{\delta m^2_j}{2m_j}
\end{equation}
The value of this effective mass is smaller than $10^{-4}$ eV for inverted
hierarchy and smaller for normal hierarchy and renders
neutrinoless double beta decay unobservable.

When L/E becomes large enough, flavor fluxes will deviate from the
canonical value of 1/3 by\cite{beacom1}

\begin{equation}
\delta P_\beta = \frac{1}{3}
\left [ \mid U_{\beta 1} \mid^2 \chi_1 \ + \mid U_{\beta 2} \mid^2
\chi_2 \ + \mid U_{\beta 3} \mid^2 \chi_3 \right ]
\end{equation}
where $\chi_i = \sin^2 (\delta m_i^2 L /4E)$

We assume that for the neutrinos from distant sources arriving
in IceCube,
$\chi_ 1 \approx 0$ but $\chi_2 = \chi_3 \approx 1/2$; i.e. $\delta
m_1^2 << \delta m^2_2$ and $\delta m^2_3$.  For example, if $\delta
m^2_1 << 10^{-17} \ eV^2$ and $\delta m^2_2, \delta m^2_3
\stackrel{\sim}{>}
10^{-15} eV^2$ then the condition for $\chi_1 \approx 0$ and $\chi_2 =
\chi_3 \approx \frac{1}{2}$ for GRB neutrinos is satisfied.

The deviation from 1/3 for $\nu_\mu's$ is given by
\begin{equation}
 \delta P_\mu   = - \frac{1}{3} \left [ \frac{1}{2} (\mid U_{\mu
      2} \mid^2 + \mid U_{\mu 3} \mid^2 ) \right ]
\end{equation}
Using the current best values for the mixing
parameters\cite{schwetz}, this   can be very close  1/6, thus giving an
extra reduction by a factor of 2 for the flux of $\nu_\mu's$.
In a model for pseudo-Dirac neutrinos via
Mirror-world, a further suppression by a factor 1/2 obtains
resulting in
a net suppression by a factor of 1/4\cite{joshipura}.
Furthermore, the shift in $P_e$ from the value 1/3 is about 0.8, and so
the ratio $\nu_e / \nu_\mu$ is about 3.

This is a very different physics possibility from the decay case
but also gives  rise to low fluxes of $\nu_\mu s$ consistent with
the lack of observation in IceCube.

To summarize, we raise two rather different possibilities
of neutrino properties which can account for the low fluxes of
$\nu_\mu's$ at high energies, and give rather large values for the
ratio of $\nu_e$ to $\nu_\mu$ fluxes. The two can
be distinguished in several
ways. The decay changes the primordial neutrino counting
from 3 to 3+4/7, and the pseudo-Dirac neutrinos make
the neutrinoless double beta decay  unobservable. The flavour ratios $\nu_e/\nu_\mu$ is another clear indicator
of the mechanism responsible; in the decay case it may vary between 2.5 and 8, whereas is 3 for the pseudo-Dirac
case.
Only further experimental data can confirm or rule out these speculations. Since the  scenarios considered here do not suppress the electron neutrino
flux, we have no problem with the PeV shower events reported by IceCube
at the Neutrino 2012 meeting in June \cite{Ishihara}.

If $\nu_\mu$ events in PeV energy range are seen in Icecube, the drastic
explanation offered here becomes unnecessary.  In that case, the
observed flavor ratios can be used to constrain parameters of models
such as the ones discussed here as has been discussed before \cite{pakvasa2}.

Most of the material in this letter was
presented by one of the authors(S.P.) at the
``What's nu-Invisibles'' workshop at GGI, Florence in
June 2012 and at the CETUP
workshop, Lead, S. D. in July  2012.

\section*{Acknowledgments:}

We thank John Learned for discussions and a careful reading of the
manuscript and we thank Danny Marfatia for useful discussions.  This work is supported in part by US DOE Grant
DE-FG02-04ER41291 and by the Indo-U.S. Science and Technology Forum
under Grant IUSSTF/JC/Physics Beyond Standard Model/23-2010/2010-2011.
We also thank the Center for Theoretical Underground Physics and Related Areas (CETUP* 2012) for its
support and hospitality. ASJ thanks the Department of Science and Technology, Government of India for support under the J. C.
Bose National Fellowship programme, grant no. SR/S2/JCB-31/2010.

\end{document}